\documentclass{aa}
\usepackage{graphicx}
\usepackage{txfonts}
\begin{document}
\title{SCUBA sub-millimeter observations of gamma-ray bursters}

\subtitle{IV. GRB 021004, 021211, 030115, 030226, 041006}

\author{I.A. Smith\inst{1}
\and R.P.J. Tilanus\inst{2}
\and N. Tanvir\inst{3}
\and V. E. Barnard\inst{2}
\and G. H. Moriarty-Schieven\inst{2}
\and D. A. Frail\inst{4}
\and R.A.M.J. Wijers\inst{5}
\and P.~Vreeswijk\inst{6}
\and E. Rol\inst{7}
\and C. Kouveliotou\inst{8,9}}

\offprints{I. Smith}

\institute{
Department of Physics and Astronomy, Rice University, 
6100 South Main, MS-108, Houston, TX 77005-1892 USA\\
\email{iansmith@rice.edu}
\and Joint Astronomy Centre, 660 N. Aohoku Place, Hilo, HI 96720 USA
\and Centre for Astrophysics Research, University of Hertfordshire, 
College Lane, Hatfield, Herts AL10 9AB, UK
\and National Radio Astronomy Observatory, P.O. Box O, Socorro, NM 87801
\and Astronomical Institute `Anton Pannekoek', 
University of Amsterdam and Center for High-Energy Astrophysics,\\
Kruislaan 403, 1098 SJ Amsterdam, The Netherlands
\and European Southern Observatory, Alonso de C\'ordova 3107, 
Casilla 19001, Santiago 19, Chile
\and Department of Physics and Astronomy, University of Leicester, 
Leicester LE1 7RH, UK
\and NASA Marshall Space Flight Center, SD-50, NSSTC, 
320 Sparkman Drive, Huntsville, AL 35805 USA
\and Universities Space Research Association}

\date{Received ; accepted }

\abstract{
We discuss our ongoing program of Target of Opportunity (ToO)
sub-millimeter observations of gamma-ray bursts (GRBs) 
using the Sub-millimetre Common-User Bolometer Array (SCUBA) 
on the James Clerk Maxwell Telescope (JCMT).
In this paper, we present the ToO observations of GRBs 
021004, 021211, 030115, 030226, and 041006.
The observations of GRBs 021004, 021211, 030226, and 041006 
all started within $\sim 1$ day of the burst, but did not detect 
any significant sub-millimeter emission from the reverse shock 
and/or afterglow.
These observations put some constraints on the models for the early
emission, although the generally poor observing conditions and/or
the faintness of these afterglows at other wavelengths limit 
the inferences that can be drawn from these lack of detections.
However, these observations demonstrate that SCUBA can perform
rapid observations of GRBs, and provide encouragement for future 
observations in the Swift era.
None of these GRBs had significant sub-millimeter emission from 
their host galaxies.
This adds to the indication that GRBs are not closely linked to 
the most luminous dusty star-forming galaxies.
}

\maketitle

\section{Introduction}

The discovery of localized transients in the error boxes of
gamma-ray burst (GRB) sources has led to intense multi-wavelength
campaigns that have revolutionized our understanding of these sources.
For reviews see Van Paradijs et al. (\cite{scuba:vanp00}) and 
M\'esz\'aros (\cite{scuba:mes02}).
The current evidence indicates that at least some GRBs are
due to the explosive collapse of massive stars.

During and after the burst, the observed multiwavelength
emission comes from several distinct components.
The explosion produces shocks that energize particles whose
radiation gives the ``prompt'' bright burst emission.
A reverse shock can give an optical and/or radio ``flash.''
The ``afterglow'' emission comes from an expanding fireball
as it sweeps up the surrounding medium.
At later times, signatures may appear that are characteristic
of supernovae or hypernovae.
Finally, the ``quiescent'' constant emission comes from any underlying 
host galaxy.

Sub-millimeter observations form a key element of
the multiwavelength observations of the bursters.
They provide ``clean'' measures of the source intensity, unaffected 
by scintillation and extinction.
To this end, we have been performing Target of Opportunity (ToO)
sub-millimeter observations of GRB counterparts using the 
Sub-millimetre Common-User Bolometer Array (SCUBA) 
on the James Clerk Maxwell Telescope (JCMT) on Mauna Kea, Hawaii.

The detailed SCUBA ToO results for the first eight bursts studied 
(GRBs 970508, 971214, 980326, 980329, 980519, 980703, 981220, and 981226) 
are described in Smith et al. (\cite{scuba:smith99});
GRB 980329 is also discussed in Yost et al. (\cite{scuba:yost02}) and 
Berger et al. (\cite{scuba:berg03}), and GRB 980703 in 
Bloom et al. (\cite{scuba:bloom98}) and Frail et al. (\cite{scuba:frail03a}).
GRB 990123 is discussed in Galama et al. (\cite{scuba:gal99nat})
and Kulkarni et al. (\cite{scuba:kfs99}).
Observations of GRB 990520 were made in mediocre weather
(Smith et al. \cite{scuba:smith00}).
The next six bursts 
(GRBs 991208, 991216, 000301C, 000630, 000911, and 000926)
are described in Smith et al. (\cite{scuba:smith01});
GRB 991216 is also discussed in Frail et al. (\cite{scuba:frail00}),
and GRB 000301C in Berger et al. (\cite{scuba:berger00}).
GRB 010222 was described in Frail et al. (\cite{scuba:frail02}).
In this paper, we present all the ToO observations of GRBs made by 
SCUBA from March 2001 through February 2005, with the exception 
of GRB 030329, which was discussed in Smith et al. (\cite{scuba:smith05}).
The results shown here are for GRBs 021004, 021211, 030115, 030226, 
and 041006.

In the remainder of \S 1, we outline the motivations for making
sub-millimeter observations of the afterglows and host galaxies and
summarize some of the previous results.
In \S 2 we describe the most relevant technical details of the SCUBA 
observations and data analysis.
In \S 3 we present the results for our latest SCUBA ToO observations.
In \S 4 we discuss the results.
In \S 5 we explain how our program will make good use of the
bursts observed by {\it Swift}.

\subsection{SCUBA afterglow observations}

Both observations (e.g. Galama et al. \cite{scuba:galama98},
\cite{scuba:gal99nat}; Smith et al. \cite{scuba:smith05}) and theories 
(e.g. Sari et al. \cite{scuba:spn98}; Piran \cite{scuba:piran99};
Wijers \& Galama \cite{scuba:wg99}; Granot et al. \cite{scuba:gps00};
Sari \& M\'esz\'aros \cite{scuba:sm00}; Chevalier \& Li \cite{scuba:cl00};
Granot \& Sari \cite{scuba:gs02}; Panaitescu \& Kumar \cite{scuba:pk04};
Inoue et al. \cite{scuba:ioc05})
show that for some bursts the reverse shock and/or afterglow emission can 
peak in the sub-millimeter in the hours to weeks following the burst.

By tracking the evolving emission across the entire spectrum, 
it is possible to study aspects such as the types of shocks involved,
the geometry of the outflow (jet versus spherical), and the geometry of
the surrounding medium (uniform versus stellar wind).
It is of interest to look for variations in the afterglow light curve
that could be due to the refreshing of the shock in the fireball 
(e.g. Sari \& M\'esz\'aros \cite{scuba:sm00};
Granot et al. \cite{scuba:gnp03}), or due to inhomogeneities 
in the ambient medium that the fireball is expanding into
(e.g. Berger et al. \cite{scuba:berger00}).

The afterglow evolution can be complex, and separating the different
components is not trivial.
Sub-millimeter observations performed within a day of the 
burst are of particular importance, since they can potentially 
strongly discriminate between different afterglow models
(Panaitescu \& Kumar \cite{scuba:pk00}; Livio \& Waxman \cite{scuba:lw00};
Yost et al. \cite{scuba:yost03}; Inoue et al. \cite{scuba:ioc05}).
Distinct early evolution behaviors could also be present in the radio
data, but they might be hard to extract due to scintillation and/or
self-absorption.
At higher frequencies, the difference in the early evolution is 
much smaller, and suffers from degeneracies.

\begin{figure}
\centering
\includegraphics[width=8.5cm]{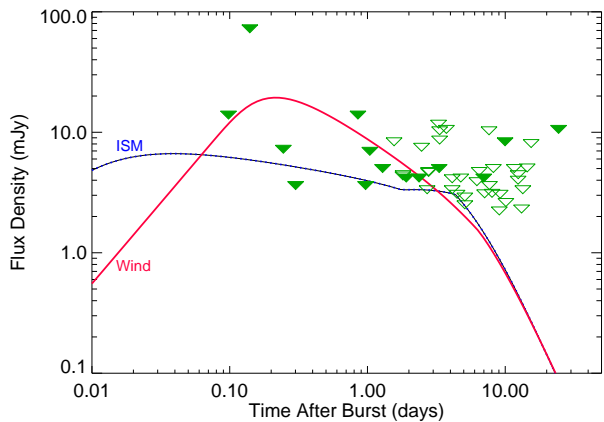}
\caption{The model curves illustrate the difference in the 850~$\mu$m 
evolution expected for a GRB that explodes in a wind-fed (Wind) or a 
constant density (ISM) circumburst medium.
The models are taken from fits made to the cm radio, near-IR, optical, 
and X-ray data of GRB 980703 and redshifted to $z = 1.5$, which is
typical for the GRBs observed to date (Frail et al. \cite{scuba:frail03a}).
The models only use a forward shock; an additional component from
a reverse shock may also be present at early times.
The $3 \sigma$ upper limits show all the 850~$\mu$m GRB ToO observations 
made by SCUBA where no afterglow was detected.
The filled triangles show the data for the new bursts presented 
in this paper.
These new observations are starting to meaningfully sample the 
sub-millimeter emission at early times where there can be
significant differences in the evolution of the sub-millimeter flux
in the wind and ISM scenarios.}
\label{figure1}
\end{figure}

The sub-millimeter flux from the forward shock in a fireball that is 
interacting with a homogeneous medium should evolve relatively slowly 
over the first few days after the burst.
However, for a fireball that is interacting with a prior stellar wind,
the forward shock is propagating down a large density gradient and the
sub-millimeter flux will rapidly rise and fall.
After a few days, the evolution of the sub-millimeter flux will be
similar for both models.
The contrasting behavior of these two scenarios is illustrated
in Figure~\ref{figure1}.

There may also be a significant sub-millimeter emission at early
times from a reverse shock.
The wavelength for the peak of the reverse shock emission and the
flux at the peak may be relatively insensitve to the redshift of 
the burst (Inoue et al. \cite{scuba:ioc05}).
Thus the reverse shock emission may be at mJy levels out to
$z \sim 30$, even if the forward shock is too faint to detect.

The most significant afterglow emission observed by SCUBA to date
was from GRB 030329 (Smith et al. \cite{scuba:smith05}).
This had a $\sim 31$~mJy flux density at 850 $\mu$m that was
essentially constant up to $\sim 7$ days after the burst.
After this it had a break to a steep decay.

Prior to GRB 030329, the most significant afterglow emission 
observed by SCUBA at 850 $\mu$m was from GRB 980329 
(Smith et al. \cite{scuba:smith99}; Yost et al. \cite{scuba:yost02}; 
Berger et al. \cite{scuba:berg03}).
The flux and rms reported by Smith et al. were higher than in the 
later papers.
This is because Smith et al. dropped the last two (out of five) runs 
from the analysis on 1998 April 5 and the third one (out of five) from 
1998 April 6.
In both cases, the declining flux during these runs suggested a pointing 
drift; on April 6, a re-pointing and re-focusing after the third run
resulted in the signal returning in the remaining runs.
Searches for a host galaxy indicate that its contribution is $< 1$ mJy 
(Tanvir et al. \cite{scuba:tan04}).
In conclusion, the fading afterglow for GRB 980329 had an 850 $\mu$m 
flux density $\sim 3$ mJy one week after the burst.

Some of the ToO observations of other bursts have produced possible
detections at 850 $\mu$m.
For GRB 990123, there was a $3.3 \sigma$ detection of 4.9 mJy on 
1999 February 4, although this was not detected in longer observations 
the following night or in earlier observations 
(Galama et al. \cite{scuba:gal99nat}; Kulkarni et al. \cite{scuba:kfs99}).
For GRB 000301C, there was an indication of a $\sim 3$ mJy source at 
850 $\mu$m, but only on 2000 March 5 (Smith et al. \cite{scuba:smith01}).
This was at the time when an achromatic brightening was found
at other wavelengths (e.g. Masetti et al. \cite{scuba:mas00}; 
Berger et al. \cite{scuba:berger00}).
The limited observation of GRB 000926 gave $7.3 \pm 4.2$ mJy at 
850 $\mu$m on 2000 September 30.
This is consistent with the multiwavelength observations of 
Harrison et al. (\cite{scuba:har01}) that showed the peak of 
the spectrum was in the sub-millimeter at that time.
Finally, the sub-millimeter emission from GRB 010222 may have
had a $\sim 1$ mJy contribution from the afterglow 
(Berger et al. \cite{scuba:berg03}).

Prior to this paper, GRB 010222 was the only case where the initial 
SCUBA observation was performed within one day of the burst.
There has therefore been limited study of the potentially bright early 
sub-millimeter emission from the afterglow and/or the reverse shock.

\subsection{SCUBA host galaxy observations}

In addition to studying the afterglows, it is important to understand 
the nature of the host galaxies containing the GRBs.

It is plausible that if GRBs are due to the explosive deaths of
high-mass stars, they are likely to be found in active star 
forming regions.
Optical studies indicate that many of the host galaxies studied to date 
are similar to young starburst galaxies with moderate to low extinction 
(e.g. Fynbo et al. \cite{scuba:fynbo03}; Christensen et al. 
\cite{scuba:christ04}).
Sub-millimeter observations can investigate whether there is a 
connection between GRBs and dust-enshrouded star formation 
(Blain \& Natarajan \cite{scuba:bn00}).
If so, estimates have suggested that $\sim 20$\% of GRB hosts will be 
brighter than 2 mJy at 850 $\mu$m (Ramirez-Ruiz et al. \cite{scuba:rtb02})
or that $7-18$\% will be brighter than 4 mJy at 850 $\mu$m 
(Trentham et al. \cite{scuba:trent02}).

SCUBA has discovered (non-GRB) sub-millimeter bright galaxies out to high 
redshifts; for a review see Blain et al. (\cite{scuba:blain02}).
It appears that the star formation rate does not drop rapidly beyond 
$z \sim 1$ (e.g. Blain et al. \cite{scuba:blain99}), and the median 
redshift for galaxies that are bright at sub-millimeter wavelengths 
is $\sim 2.2$ (Chapman et al. \cite{scuba:chap03}, \cite{scuba:chap05}).
Many of the sub-millimeter bright sources may be undergoing major
mergers (Conselice et al. \cite{scuba:con03}).

The relatively small population of extremely luminous dusty
galaxies may dominate the total star formation in the universe at
early epochs.
However, a difficulty in reaching this conclusion is that
there may be a substantial contribution to the energy output 
of the SCUBA-bright galaxies from active galactic nuclei.
Furthermore, some galaxies are sub-millimeter bright simply due to 
gravitational lensing (e.g. Dunlop et al. \cite{scuba:dun04}).
If GRBs are regularly found to be associated with extremely luminous 
sub-millimeter bright dusty galaxies, this could provide independent 
evidence that these objects dominate the star formation in the early 
universe.
The gamma-rays from the bursts can be detected out to high redshift, 
and through large columns of dust and gas.

Our ToO program is designed to look for the afterglow emission of the 
bursts by making observations shortly after the burst, with follow-up 
observations on subsequent days.
Since these ToO observations are done on short notice, their sensitivity 
can be limited by non-optimal weather and/or source location in the sky.
However, by combining the data from all of our ToO observations of a 
source we can often put useful constraints on the quiescent emission
(Smith et al. \cite{scuba:smith01}).

Complementary programs have used SCUBA to perform sensitive searches 
for the quiescent emission from GRB hosts long after any afterglow 
emission has faded away (Barnard et al. \cite{scuba:barn03}; 
Berger et al. \cite{scuba:berg03}; Tanvir et al. \cite{scuba:tan04}).
These observations are generally done in good weather conditions
when the source is at a high elevation.

To date, of the 21 GRB host galaxies with 850 $\mu$m rms values 
$< 1.4$ mJy (not including the new observations described here), 
none have had a flux density $> 4$ mJy.
The only host galaxies significantly detected ($> 3 \sigma$) by SCUBA
to date are GRBs 000210, 000418, and 010222.
The tentative dearth of brighter hosts suggests that the GRBs do not 
in general trace the most luminous sub-millimeter galaxies.
However, it should be cautioned that the current sample of bursts
is small.
For example, only 6 of the 21 galaxies with measured redshifts
in Table 2 of Tanvir et al. (\cite{scuba:tan04}) have $z>2$.
There also could be important selection effects since many of the 
bursts studied by SCUBA have had optical transients and thus may have 
relatively little absorption local to the source.
It is also possible that the dust temperatures tend to be higher in 
GRB host galaxies, shifting the peak of their dust emission to shorter
wavelengths and reducing the sub-millimeter flux below the current
sensitivity of SCUBA 
(Chapman et al. \cite{scuba:chap04}).

In Smith et al. (\cite{scuba:smith01}) we gave numerical estimates 
for the expected sub-millimeter emission from warm dust due to massive 
star formation using the formalism from Condon (\cite{scuba:condon})
and Carilli \& Yun (\cite{scuba:cy99}).
[For completeness, we can now fill in the entry for GRB 000911 in Table 2 
of Smith et al. (\cite{scuba:smith01}) using $z=1.0585$ 
(Price et al. \cite{scuba:price02}) to give a 
${\rm SFR}~ (M \ge 5 M_{\sun}) < 220 ~M_{\sun} {\rm yr}^{-1}$.]
However, this conversion currently has very large uncertainties due to 
the lack of information for key quantities such as the temperature of 
the dust, the spectral indices in the sub-millimeter region and the 
initial mass function.
For example, for a given 850 $\mu$m flux density, doubling the dust 
temperature corresponds to changing the inferred luminosity by a 
factor of $\sim 10$ (Blain et al. \cite{scuba:blain02}).
We therefore simply quote here the flux densities that we find for the 
new bursts.
These can be directly compared to the previous results
(Smith et al. \cite{scuba:smith01}, \cite{scuba:smith05}; 
Barnard et al. \cite{scuba:barn03}; Berger et al. \cite{scuba:berg03}; 
Tanvir et al. \cite{scuba:tan04}).

In the future, it may be possible to combine the sub-millimeter
observations with those at mid- and far-IR wavelengths, e.g. using
the {\it Spitzer Space Telescope}.
This combination -- along with the known burst redshift -- will allow a
good modeling of the spectral energy distribution, from which the
temperature and luminosity of the dust can then be determined
(Blain et al. \cite{scuba:blain03}).
The star formation rate can then be calculated to a much higher
accuracy than using the sub-millimeter observations alone.

\section{SCUBA observing details}

SCUBA is the sub-millimeter continuum instrument for the JCMT
(for a review see Holland et al. \cite{scuba:hol99}).
Here we summarize only the most important features of the instrument.
The observing, calibration, and reduction techniques are the same as 
in Smith et al. (\cite{scuba:smith99,scuba:smith01}).

SCUBA uses two arrays of bolometers to simultaneously observe the same
region of sky, $\sim 2.3\arcmin$ in diameter.
The arrays are optimized for operations at 450 and 850 $\mu$m.
Fully sampled maps of the $2.3\arcmin$ region can be made by ``jiggling'' 
the array.
However, for all the sources described here, we have only been looking at 
well-localized optical or radio transient coordinates by performing deep
photometry observations using a single pixel of the arrays.
The other bolometers in the arrays are used to perform a good sky noise 
subtraction (Archibald et al. \cite{scuba:arch02}).

During an observation, the secondary is chopped between the
source and sky.
This is done mainly to take out small relative DC drifts between the
bolometers, and also to remove any large-scale sky variations.
The term ``integration time'' always refers to the ``on+off''
time, including the amount of time spent off-source.
An 18 sec integration thus amounts to a 9 sec on-source observation time.
A typical measurement consists of 50 integrations of 18 seconds;
we refer to this as a ``run.''
Each observation of a source in general consists of several such
runs, with focus, pointing, and calibration observations in between.

Version 1.6 of SURF\footnote{SURF is distributed by the Starlink Project.}
(Jenness \& Lightfoot \cite{scuba:jen00}) was used to combine the 
integrations, remove anomalous spikes, flatfield the array, and 
subtract the signal from the sky bolometers.
The zenith sky opacity was determined using the JCMT Water Vapor Monitor.  
This is located in the Receiver Cabin, and is pointed just slightly 
off the main beam.  
It measures the opacity every 6 seconds.  
The results were checked using ``skydips'' in which the sky brightness 
temperature was measured at a range of elevations.
The opacity was used to apply atmospheric extinction corrections to the 
observed target fluxes.
At least one standard calibration target was observed each night to 
determine the absolute flux of the GRB.

A typical integration time of 2 hours gives an rms $\sim 1.5$~mJy at 
850 $\mu$m.
However, the sensitivity depends significantly on the weather and
the elevation of the source; since our ToO observations are done
on short notice, sometimes these factors are less than ideal.
Based on observed variations of the gain factor and signal levels we 
estimate typical systematic uncertainties in the absolute flux calibrations 
of 10\% at 850 $\mu$m.
In general the rms errors of the observations presented here are larger than 
this uncertainty.

\subsection{Pointing errors}

The pointing of the JCMT is checked several times during the night 
to ensure that it is reliable.
The pointing accuracy is usually a few arcsec.
However, an error in the track model used between 2000 August 25 and
2003 April 25 resulted in pointing errors that were non-negligible
(shifts larger than $4\arcsec$) for targets with elevations above 
$60\degr$ and over small ranges of azimuth.
Since the 850 $\mu$m bolometric pixel has a diffraction limited resolution
of $14\arcsec$, the target remains well within the bolometric pixel.
However, there will be an error in the flux that is measured.
This issue can be a problem if either the source or a pointing
or calibration target was affected.

For the new bursts reported here, there were small pointing problems
in some of the runs for GRBs 021004 and 030226.
These instances are noted in \S 3.
None of them significantly affected the final results.

For completeness, we note that the pointing problem also affected 
the previously reported ToO observations of GRB 000911 
(Smith et al. \cite{scuba:smith01}).
One of the four runs on 2000 September 17 was affected.
Even if we drop that day entirely, the 850 $\mu$m flux density of
$0.3 \pm 1.1$ mJy measured on 2000 September 20 is still marginally 
inconsistent with the value of $2.31 \pm 0.91$ mJy measured by 
Berger et al. (\cite{scuba:berg03}) in their quiescent host program.
[The observation on 2000 September 22 was also affected by the 
pointing problem, but this was not on the correct coordinates 
(Smith et al. \cite{scuba:smith01}).]

\subsection{False positives}

The large beam size combined with the large number of distant galaxies 
radiating strongly at this wavelength means that in any observation 
there is a non-negligible chance of detecting a quiescent sub-millimeter 
source that is completely unrelated to the GRB.

The surface density of sub-millimeter galaxies is still somewhat 
uncertain (Blain et al. \cite{scuba:blain02}).
At 850 $\mu$m, the surface density of galaxies with flux densities larger 
than 4 mJy is $\sim 10^3~{\rm deg}^{-2}$, while the surface density of 
galaxies with flux densities larger than 1 mJy is $\sim 10^4~{\rm deg}^{-2}$.
This gives estimates that the chance of detecting a random $\ge 4$ mJy 
source in any pointing is $\sim 1$\%, while the chance of detecting a 
random $\ge 1$ mJy source in any pointing is $\sim 10$\%.

The non-negligible chance of a false positive means that caution
is required when using just the detection of a quiescent sub-millimeter 
source to claim that this must be the host galaxy to the GRB.
For individual cases, confirmation of the star formation rate is
needed from observations at other wavelengths.
However, if it is found that many more bursts are associated with 
quiescent sub-millimeter sources than is expected by chance, this would 
be good evidence that the majority of these are true associations.

\begin{table*}
\begin{minipage}[t]{\columnwidth}
\caption{SCUBA 850~$\mu$m (350 GHz) GRB afterglow observations.}
\label{table:all850}      
\centering 
\renewcommand{\footnoterule}{}
\begin{tabular}{l l l l l l l l}
\hline\hline       
Burst       & \multicolumn{2}{c}{SCUBA observing times} & Time since   & Integration & $\tau_{850}$ & Afterglow 850 $\mu$m \\
            & Start                & Stop               & burst (days) & time (sec)  &              & flux density (mJy)   \\
\hline                    
GRB 021004  & 20021004.622 & 20021004.665 & \phantom{1}0.140 & 1980 & 0.47 & $\phantom{-4.5} 5 \pm 26$ \\
            & 20021005.349 & 20021005.365 & \phantom{1}0.853 & \phantom{1}900 & 0.44 & \phantom{1}$-4.0 \pm 5.0$ \\
            & 20021006.353 & 20021006.481 & \phantom{1}1.913 & 4662 & 0.27 & $\phantom{-1} 0.9 \pm 1.5$ \\
            & 20021011.434 & 20021011.543 & \phantom{1}6.984 & 5400 & 0.19 & \phantom{1}$-3.7 \pm 1.5$ \footnote{Not corrected for small pointing error} \\
\\
GRB 021211  & 20021212.488 & 20021212.520 & \phantom{1}1.033 & 1800 & 0.23 & $\phantom{-1} 2.1 \pm 2.5$ \\
            & 20021221.429 & 20021221.460 & \phantom{1}9.973 & 1800 & 0.40 & \phantom{1}$-2.4 \pm 3.0$ \\
\\
GRB 030115  & 20030118.410 & 20030118.501 & \phantom{1}3.314 & 4500 & 0.34 & $\phantom{-1} 0.2 \pm 1.8$ \\
\\
GRB 030226  & 20030226.386 & 20030226.418 & \phantom{1}0.245 & 1800 & 0.35 & $\phantom{-1} 5.1 \pm 2.6$ \\ 
            & 20030226.386 & 20030226.531 & \phantom{1}0.301 & 6498 & 0.35 & $\phantom{-1} 1.2 \pm 1.3$ \\
            & 20030227.412 & 20030227.477 & \phantom{1}1.288 & 3600 & 0.35 & \phantom{1}$-1.9 \pm 1.8$ $^{a}$ \\
            & 20030228.489 & 20030228.564 & \phantom{1}2.369 & 3618 & 0.22 & $\phantom{-1}0.0 \pm 1.5$ $^{a}$ \\
            & 20030322.542 & 20030322.574 &           24.401 & 1800 & 0.48 & \phantom{1}$-4.3 \pm 3.8$ \\
\\
GRB 041006  & 20041006.604 & 20041006.618 & \phantom{1}0.098 & \phantom{1}720 & 0.25 & $\phantom{-1} 2.0 \pm 5.0$ \\
            & 20041007.374 & 20041007.590 & \phantom{1}0.969 & 4320 & 0.23 & $\phantom{-1} 0.9 \pm 1.3$ \\
\hline 
\end{tabular}
\end{minipage}
\end{table*}

\section{Results for new SCUBA observations}

Table~\ref{table:all850} summarizes all the 850 $\mu$m SCUBA 
ToO observations for the new bursts.
The start times are the times when the GRB was first observed on
each day, and the stop times are when the last observation of the 
GRB was completed.
The time since the burst uses the mid-point between these start and
stop times (although it should be noted that focus, pointing, and 
calibration observations may have also taken place in between the 
start and stop times); the time is the elapsed time in the Earth 
frame, not the rest frame of the host galaxy.
The ``integration time'' for each observation is the ``on+off'' time; 
only half of this is spent on-source.
The zenith optical depth at 850 $\mu$m ($\tau_{850}$) is given for
the time of the observation.

The $3 \sigma$ upper limits for all the new ToO observations 
are shown as the filled triangles in Figure~\ref{figure1}.

For each burst, the 850~$\mu$m data from the separate days (that have 
no obvious afterglow emission) was combined using a weighted mean to 
give the most sensitive value for the flux of the quiescent host galaxy.
These results are given in Table~\ref{table:host}.
Our observations are close enough to the time of the burst that a 
small amount of afterglow emission may be present in the final results.
However, this contamination will be small.

In the following sections, we discuss each of these bursts in
more detail.

\subsection{GRB 021004}

The long ($\sim 100$ sec) GRB 021004 was detected by {\it HETE-2} at
20021004.504 UT (Shirasaki et al. \cite{scuba:shir02}).
The burst was X-ray rich (Lamb et al. \cite{scuba:lamb02}).
A fading X-ray source was found by the {\it Chandra X-Ray Observatory}
whose light curve decay was variable, but whose spectrum showed no 
significant features (Fox et al. \cite{scuba:fox03a}; 
Butler et al. \cite{scuba:but03}; Holland et al. \cite{scuba:hol03}).

Observations starting 193 seconds after the burst detected an 
$R \sim 15.5$ optical afterglow (Fox et al. \cite{scuba:fox03a}).
The large number of lines in the optical spectrum complicated the 
determination of the redshift of the counterpart, but most likely it 
was an active starburst $R \sim 24$ galaxy at $z = 2.328$ 
(e.g. Mirabal et al. \cite{scuba:mir03}; Schaefer et al. \cite{scuba:sgh03}).

Unlike GRBs 990123 and 021211, where the optical light curve had a 
rapid initial fade before flattening to a value typical of afterglows,
for GRB 021004 the afterglow had an initial decay for the first $\sim 20$
minutes but then a plateau that lasted for $\sim 3$ hours
(Uemura et al. \cite{scuba:ukiy03}).
There were significant variations on several time scales superposed on 
the later optical decay, and there were significant color and
polarization changes during the afterglow
(e.g. Bersier et al. \cite{scuba:ber03}; Matheson et al. \cite{scuba:mat03};
Rol et al. \cite{scuba:rol03}; Lazzati et al. \cite{scuba:laz03}).
This suggests that there could have been repeated energy injections
from the central source (e.g. Bj\"ornsson et al. \cite{scuba:bgj04}), 
and/or the surrounding medium may have been significantly inhomogeneous.
Highly ionized lines with high relative velocities were seen in the 
spectrum; these may have come from shells or clumps that could have 
been produced by a massive stellar progenitor prior to its collapse
(e.g. Mirabal et al. \cite{scuba:mir03}; Schaefer et al. \cite{scuba:sgh03}).
On larger scales, the ambient medium may have been homogeneous.
A break in the optical light curve $\sim 5 - 9$ days after the 
burst was most likely the jet break
(Holland et al. \cite{scuba:hol03}; Mirabal et al. \cite{scuba:mir03}).

A variable radio and millimeter counterpart was found 
(Frail \& Berger \cite{scuba:fb02}; Pooley \cite{scuba:pool02};
Bremer \& Castro-Tirado \cite{scuba:bct02}).
The longer wavelength data was fitted by a simple power law
$S_\nu \propto \nu^{\alpha}$ with $\alpha \sim 0.9$.
This spectral index was considered to be unusual 
(Berger et al. \cite{scuba:berg02}),
although a similar result had previously been discussed for
GRB 980329 (Smith et al. \cite{scuba:smith99}).
At $\sim 1.4$ days after the burst, the peak of the multiwavelength 
spectrum was $\sim 3$~mJy somewhere in the range $100 - 1000$~GHz
(Schaefer et al. \cite{scuba:sgh03}).

All the SCUBA 850~$\mu$m observations of GRB 021004 are shown in 
Table~\ref{table:all850}.
These expand on those given previously (Kemp et al. \cite{scuba:kemp02};
Wouterloot et al. \cite{scuba:wout02}).
On 2002 October 11 there were small pointing problems (see \S 2.1) with 
two of the SCUBA runs on the pointing calibration target PKS~0106+013.
These runs do not significantly affect the results for that day.

Our first SCUBA observation of GRB 021004 started less than 3 hours 
after the burst.
However, the source was at a low elevation and was setting, and the 
weather was only fair.
The upper limit was therefore much worse than normal.
A second observation at a good elevation was attempted later the same 
day, but this had to be cut short due to fog at the summit.
Again, the upper limit that was obtained was poor.
Better observations were made on following days.
The source was not detected by SCUBA in any of the observations.

Our non-detections are consistent with the lack of detection 
($1.9 \pm 1.5$~mJy at 1.5 days after the burst) by the IRAM Plateau de 
Bure Interferometer at 232 GHz (Bremer \& Castro-Tirado \cite{scuba:bct02}).
At that time, the peak of the afterglow spectrum was only $\sim 3$~mJy.
Our results are also consistent with the wind-interaction model of 
Li \& Chevalier (\cite{scuba:lc03}) that placed the peak of the
350 GHz flux at $\sim 8$~mJy around $\sim 0.3$ days after the burst.

As shown in Table~\ref{table:host}, the combination of our ToO observations 
puts good limits on the quiescent sub-millimeter flux of the host galaxy.
These results improve slightly on those shown in 
Tanvir et al. (\cite{scuba:tan04}).
The unobscured star formation rate estimated from the Ly$\alpha$ 
luminosity is $\sim 15 M_{\sun}$/year 
(Djorgovski et al. \cite{scuba:djo02}).
Although this is an active starburst galaxy whose redshift is
approximately at the peak for the sub-millimeter bright galaxies,
our limit implies that there is not a substantial amount of
obscured star formation in the host galaxy.
The lack of detection of a sub-millimeter host galaxy agrees with the 
X-ray observations of the afterglow that show no evidence for absorption 
in excess of the Galactic value (Fox et al. \cite{scuba:fox03a}; 
Butler et al. \cite{scuba:but03}; Holland et al. \cite{scuba:hol03}).
Some reddening due to local extinction in the host galaxy is required 
to explain the optical spectrum, but this is not substantial in the 
case of a circumstellar medium with a wind-like density profile.

\begin{table}[t]
\caption{Host galaxy fluxes for each burst determined by combining the 
SCUBA 850~$\mu$m observations taken on separate days.}
\label{table:host}
\centering
\begin{tabular}{l r l}
\hline\hline
Burst       & Redshift     & Host 850 $\mu$m flux \\
            &              & density (mJy)        \\
\hline
GRB 021004  & 2.328        & $           -1.4 \pm 1.0$ \\
GRB 021211  & 1.006        & $\phantom{-} 0.3 \pm 1.9$ \\
GRB 030115  & $2-2.5$      & $\phantom{-} 0.2 \pm 1.8$ \\
GRB 030226  & $\geq 1.986$ & $           -0.1 \pm 0.8$ \\
GRB 041006  & 0.716        & $\phantom{-} 1.0 \pm 1.3$ \\
\hline
\end{tabular}
\end{table}

\subsection{GRB 021211}

The $\sim 10$ second long FRED-like GRB 021211 was a bright, 
X-ray rich burst detected by {\it HETE-2} at 20021211.471 UT 
(Crew et al. \cite{scuba:crew03}).

Rapid optical observations detected an afterglow
that faded quickly from $R \sim 14$ at 90 seconds after the burst to 
$R \sim 18$ after 20 minutes (Wozniak et al. \cite{scuba:woz02};
Park et al. \cite{scuba:pwb02}; Fox et al. \cite{scuba:fox03b};
Li et al. \cite{scuba:li03}).
The initial decay rate was steep, but after $\sim 10$ minutes
this slowed to a value more typical of optical afterglows.
The initial emission may have come from the reverse shock.
Unlike GRB 021004, the optical afterglow for GRB 021211 resembled the one 
for GRB 990123 at similar epochs, but it was $\sim 3-4$ magnitudes fainter.
If it had not been for the early observations, the burst might have 
been classified as ``optically dark.''

There appear to have been some short time scale variations superposed
on the overall optical light curve decay, but there was only very weak 
evidence for a jet break (Holland et al. \cite{scuba:hol04}).
A rebrightening in the optical light curve $\sim 25$ days after the burst 
may have been due to a supernova (Della Valle et al. \cite{scuba:dv03}).
An underlying $R \sim 25.2$ host galaxy was found 
(Della Valle et al. \cite{scuba:dv03}).
The redshift of the host was $z = 1.006$ 
(Vreeswijk et al. \cite{scuba:vrees03}).

The source was not significantly detected in any of the radio 
observations that were made, putting constraints on the reverse shock
(Rol \& Strom \cite{scuba:rs02}; Fox et al. \cite{scuba:fox03b}).

All the SCUBA 850~$\mu$m observations of GRB 021211 are shown in 
Table~\ref{table:all850}.
These expand on those given previously (Hoge et al. \cite{scuba:hoge02};
Fox et al. \cite{scuba:fox03b}).
Our first observation was made $\sim 1$ day after the burst.
The source was not detected by SCUBA in either observation.

If we extrapolate from the optical spectrum using the simple 
synchrotron model $S_\nu \propto \nu^{- \beta}$ with $\beta = 0.6$ 
as in Pandey et al. (\cite{scuba:pan03}), the expected 850~$\mu$m 
flux density would have been $\sim 0.2$ mJy at 1 day after the burst.
Similarly, using $\beta = 0.69$ as in Holland et al. (\cite{scuba:hol04})
would give $\sim 0.3$ mJy.
Thus although GRB 021211 had a detectable optical reverse shock, it is
not surprising that it was not detected in the sub-millimeter
given the faintness of the optical afterglow at the times the SCUBA 
observations were made, and the lack of radio detection at any time.

As shown in Table~\ref{table:host}, the combination of our ToO 
observations limits the quiescent sub-millimeter flux of the host galaxy.
Given the relatively low redshift, and the lack of extinction in
the host galaxy (Holland et al. \cite{scuba:hol04}), it is not 
surprising that it is not a bright sub-millimeter source.
The fact that no SCUBA source was detected confirms that this burst 
had an intrinsically faint afterglow, rather than one that was heavily 
absorbed by dust in the host galaxy.

\subsection{GRB 030115}

The $\sim 20$ second long GRB 030115 was detected by {\it HETE-2} at 
20030115.141 UT (Kawai et al. \cite{scuba:kawai03}).
Optical observations simultaneous with the burst did not detect the 
source, with an upper limit of $R \sim 10$ 
(Castro-Tirado et al. \cite{scuba:ct03}).
Early optical afterglow searches did not detect a counterpart, but a 
fading infrared source was found (Levan et al. \cite{scuba:levan03};
Vrba et al. \cite{scuba:vrba03}; Masetti et al. \cite{scuba:mas03};
Kato et al. \cite{scuba:kato03}).
A possible $R = 25.2$ host galaxy was found (Garnavich \cite{scuba:gar03};
Dullighan et al. \cite{scuba:dull04}).
Although no spectroscopic redshift has been reported to date,
the photometric redshift lies in the range $z=2-2.5$
(Levan et al. \cite{scuba:levan05}).
A faint ($< 0.1$ mJy) radio source was detected at the afterglow location
(Frail \& Berger \cite{scuba:fb03}; Rol \& Wijers \cite{scuba:rw03}).

Since the afterglow was faint, and there were limited observations
being made at other wavelengths, we only performed one SCUBA observation
of GRB 030115.
As shown in Table~\ref{table:all850} -- which updates the result of 
Hoge et al. (\cite{scuba:hoge03})
-- the source was not detected at 850 $\mu$m.
This is consistent with an observation made by MAMBO on IRAM a few hours
earlier that found $2.9 \pm 1.6$ mJy at 1.2 mm
(Bertoldi et al. \cite{scuba:bert03}).

Whilst the red $R-K$ color of the afterglow and host galaxy indicates 
significant reddening (Levan et al. \cite{scuba:levan05}), the 
non-detection by SCUBA, albeit with a fairly large uncertainty, 
suggests that it is not a massive star-bursting, dusty galaxy.

\subsection{GRB 030226}

The long ($> 100$ sec) GRB 030226 was detected by {\it HETE-2} at
20030226.157 UT (Suzuki et al. \cite{scuba:suz03}).
Optical observations simultaneous with the burst did not detect the 
source, placing an upper limit of $R = 11.5$ on the reverse shock emission
(Klose et al. \cite{scuba:klose04}).
A fading optical counterpart was found 2.6 hours after the burst
(Fox et al. \cite{scuba:fox03c}; Price et al. \cite{scuba:price03a}), 
and a fading X-ray source was found by the {\it Chandra X-Ray Observatory}
(Pedersen et al. \cite{scuba:ped03}; Klose et al. \cite{scuba:klose04}).
The optical afterglow started faint and faded quickly, with
an achromatic break $\sim 0.8$ days after the burst indicating 
a jetted explosion (Klose et al. \cite{scuba:klose04}).
The redshift of the optical counterpart was determined to be $\geq 1.986$ 
(Ando et al. \cite{scuba:ando03}; Price et al. \cite{scuba:price03b};
Chornock \& Filippenko \cite{scuba:cf03}; Klose et al. \cite{scuba:klose04}).
No optical host galaxy was detected down to a limit of $R = 26$
(Klose et al. \cite{scuba:klose04}).

As shown in Table~\ref{table:all850}, our first observation of GRB 030226 
with SCUBA started 5.5 hours after the burst.
We give two results for the observations on the first night.
The first 100 integrations indicated the presence of a source
at the $\sim 2 \sigma$ level.
However, the complete data set for that night did not confirm
a significant detection, possibly because the sub-millimeter
flux was already falling.
The observations on following nights did not detect the source.

There was a small pointing problem (see \S 2.1) with one of the four 
SCUBA runs on 2003 February 27.
On 2003 February 28, there was a problem with a pointing calibration
observation of 1156+295.
These runs do not significantly affect the results for those days.
There were no problems on 2003 February 26 that could explain the
possibly varying source.

Millimeter observations with the Plateau de Bure Interferometer
did not detect the source (Pandey et al. \cite{scuba:pan04}),
but these did not start until 1.8 days after the burst.
From the radio catalog of Frail et al. (\cite{scuba:frail03b}),
VLA observations 0.12 days after the burst did not detect the source,
with a flux density of $-0.058 \pm 0.074$ mJy at 8.46 GHz.
A $\sim 0.1$ mJy source was detected at 8.46 GHz starting at 1.09 
days after the burst.
Ryle observations 0.94 days after the burst gave $0.19 \pm 0.22$ mJy 
at 15 GHz, and there was no detection with OVRO 2.27 days after the 
burst at 98 GHz.
Given the sparse longer wavelength coverage and larger rms of the first 
VLA observation, there is limited information regarding the reality of 
the possible SCUBA source at early times.

If we extrapolate from the optical spectrum using the 
simple synchrotron model $S_\nu \propto \nu^{- \beta}$ with 
$\beta = 0.55$ as in Figure 2 of Pandey et al. (\cite{scuba:pan04}),
the expected 850~$\mu$m flux density would have been $\sim 4$ mJy
at 0.62 days after the burst.
A similar flux is expected if we instead extrapolate using 
$\beta = 0.7$ at the time of the optical light curve break as in 
Figure 8 of (Klose et al. \cite{scuba:klose04}).
Thus our observations -- that were made earlier than these times when 
the optical afterglow was brighter -- appear to lie a little below these
extrapolations.

Klose et al. (\cite{scuba:klose04}) noted that the optical/infrared
light curves had some deviations from the best fit model.
For example, the $K$-band flux was 0.4 mag above the best fit model 
at 0.18 days after the burst.
Although the fluctuations had a lower amplitude than in other bursts, 
they might indicate expanding shells around a massive Wolf-Rayet star. 
The optical/infrared light curves were very sparsely sampled at early 
times, but suggest that deviations were not present at the time of our 
first SCUBA observation.
Thus it is not clear if our initial SCUBA observations can be explained 
as being due to a short-lived fluctuation.
If we assume that there was an early sub-millimeter source that
faded quickly, this would favor the scenario where the fireball was
interacting with a stellar wind rather than a homogeneous medium.
This would be consistent with the suggestion of a Wolf-Rayet progenitor.

As shown in Table~\ref{table:host}, the combination of our ToO observations 
puts good limits on the quiescent 850~$\mu$m flux of the host galaxy.
Similarly, at 450~$\mu$m the combined flux density is $1 \pm 10$~mJy.
Although the redshift of the galaxy is at approximately the peak 
for sub-millimeter bright galaxies, the host for GRB 030226 is not 
a bright dusty galaxy.
This agrees with the broadband optical spectrum, which shows no 
evidence for additional reddening by dust in the host galaxy, and 
the low polarization of the afterglow, which argues against substantial 
dust extinction (Klose et al. \cite{scuba:klose04}).

\subsection{GRB 041006}

The $\sim 25$ second long GRB 041006 was detected by {\it HETE-2} at
20041006.513 UT (Galassi et al. \cite{scuba:gal04}).
The burst was reminiscent of GRB 030329 in shape and spectral properties, 
but was 20 times fainter.
A fading X-ray source was detected by {\it Chandra}
(Butler et al. \cite{scuba:but04}).
A fading $\sim 17$th magnitude optical counterpart was found starting a
few minutes after the burst 
(e.g. da Costa, Noel, \& Price \cite{scuba:dcnp04}; 
Price, da Costa, \& Noel \cite{scuba:pdcn04a}, \cite{scuba:pdcn04b}; 
Maeno et al. \cite{scuba:mae04}). 
There was a break in the afterglow light curve $\sim 7$ hours after the burst
(Kahharov et al. \cite{scuba:kah04}; D'Avanzo et al. \cite{scuba:dav04}).
The break was consistent with being achromatic, suggesting the 
explosion was jetted.
The redshift of the optical counterpart was determined to be $0.716$ 
(Fugazza et al. \cite{scuba:fug04}; Price et al. \cite{scuba:prr04}).
A host galaxy with $R \ga 25$ was found (Fynbo et al. \cite{scuba:fyn04};
Covino et al. \cite{scuba:cov04}).
Radio observations with the VLA starting at 0.74 days after the burst did 
not detect a source at 4.86 GHz or 8.46 GHz, with rms values of 0.059 mJy 
and 0.041 mJy respectively (Soderberg \& Frail \cite{scuba:sod04}).

As shown in Table~\ref{table:all850}, our first observation of GRB 041006 
with SCUBA started 2.2 hours after the burst.
The source was not detected (Barnard et al. \cite{scuba:bar04a}).
Unfortunately, the source was setting, so it was not possible to 
perform a longer observation to get a lower rms or to look for
any evolution in the source flux.
Longer observations the following night also did not detect the 
source (Barnard et al. \cite{scuba:bar04b}).

The 850~$\mu$m flux density for GRB 030329 was $\sim 31$ mJy
during the first week.
Therefore, our observations of GRB 041006 are consistent with this 
source being similar to GRB 030329, but with a flux that is 20 
times lower.

As shown in Table~\ref{table:host}, the combination of our ToO 
observations gives a good limit to the quiescent sub-millimeter 
flux of the host galaxy.
Given the relatively low redshift, it is not surprising that it is 
not a bright sub-millimeter source.

\section{Discussion}

The observations of GRBs 021004, 021211, 030226, and 041006 
all started within $\sim 1$ day of the burst, but did not detect 
any significant sub-millimeter emission from the reverse shock 
and/or afterglow.
These observations put some constraints on the models for the early
emission, although the generally poor observing conditions and/or
the faintness of these afterglows at other wavelengths limit 
the inferences that can be drawn from these lack of detections.
However, Figure~\ref{figure1} shows that we have begun to meaningfully 
sample the sub-millimeter emission at early times where there can be
significant differences in the evolution of the sub-millimeter flux
in the wind and ISM scenarios.

None of the new GRBs studied here had a significant sub-millimeter 
emission from the host galaxy.
In particular, the host galaxies for GRBs 030226 and 041006 had 
850~$\mu$m rms values $< 1.4$ mJy.
These can be added to the compilation in Table 2 of 
Tanvir et al. (\cite{scuba:tan04}) to give 23 host galaxies with
rms values $< 1.4$ mJy.
The redshifts of GRBs 030115 and 030226 were $\ga 2$.
Adding all the bursts presented here and GRB 030329
(Smith et al. \cite{scuba:smith05}) to Table 2 of 
Tanvir et al. (\cite{scuba:tan04}), there are now
8/26 host galaxies with measured redshifts $\ga 2$.
The lack of new detections adds to the indication that GRBs are not 
closely linked to the most luminous dusty star-forming galaxies.
However, as was pointed out in \S 1.2, the sample size remains small.
There is also the possible selection effect that this may not be a 
representative sample of the whole host galaxy population because 
many of the GRBs studied by SCUBA to date have had optical transients.
It is also possible that GRB host galaxies tend to have hotter dust,
shifting the peak of their far-IR emission to shorter wavelengths.

Observations of new bursts are continuing to produce surprises, and 
there is much left to learn about GRB afterglows and host galaxies. 
To obtain a complete picture of their nature will require the careful 
study of many bursts to expand our sample.
Sub-millimeter observations with a $\sim$ mJy sensitivity are a key 
component to the multi-wavelength coverage.
To this end, our program of ToO observations using SCUBA is ongoing.

\section{The future with Swift}

Only $\sim 3$ bursts per year have been observed by the SCUBA ToO 
program over the past 7 years.
This has been due to a combination of (1) the lack of well-localized 
sources in regions of the sky accessible to SCUBA, (2) downtimes to SCUBA, 
and (3) poor weather over this period.

This situation should improve significantly over the coming years 
because of the rapid burst location capabilities of {\it Swift} 
(Gehrels et al. \cite{scuba:swift04}).
{\it Swift} should localize $\sim 100$ bursts per year
(its gamma-ray instrument will be the most sensitive GRB
detector flown to date, and so the number of bursts it
will detect depends on the uncertain number of faint sources).
This will allow us to focus our SCUBA observations on the bursts that 
appear to be the most interesting and those that occur in good observing 
conditions.

Using its gamma-ray imaging capabilities, {\it Swift} will distribute 
$\sim 4\arcmin$ localizations within 8 seconds.
The satellite will then automatically repoint so that
the source is in the field of view of the on-board X-ray
and UV-optical instruments.
If an X-ray counterpart is present, coordinates with an accuracy of
$\sim 5\arcsec$ will be available $\sim 96$ seconds after the burst.
If an optical counterpart is present, coordinates with an accuracy of
$\sim 0.3\arcsec$ will be available $\sim 243$ seconds after the burst.
Even if only an X-ray source is present, we will be able to immediately 
use SCUBA in the photometry mode (the 850 $\mu$m bolometric pixel has 
a diffraction limited resolution of $14\arcsec$).
Thus we will be able to look for sources that may be hard to detect 
otherwise if the redshift is large and/or if the optical extinction 
is large.

It is also exciting that {\it Swift} will be able to localize bursts that
last less than 1 second: these may have different progenitors and 
counterpart behaviors from the objects studied to date 
(Kouveliotou et al. \cite{scuba:kouv93};
Panaitescu, Kumar, \& Narayan \cite{scuba:pkn01}).
All of the bursts studied by SCUBA to date have been in the 
long duration class.

The results presented here demonstrate that we can perform 
observations with SCUBA shortly after a burst is reported.
The possible ($\sim 2 \sigma$) source in the early observation of
GRB 030226 provides encouragement for future similar observations
to study the reverse shock and early fluctuations in the afterglow
light curve.
We are therefore in an excellent position to take advantage of Swift.

\begin{acknowledgements}

The James Clerk Maxwell Telescope is operated by The Joint Astronomy 
Centre on behalf of the Particle Physics and Astronomy Research Council 
of the United Kingdom, the Netherlands Organisation for Scientific Research, 
and the National Research Council of Canada.

We thank the JCMT Directors Ian Robson and Gary Davis for authorizing 
the ToO observations.
We are indebted to all the observers whose time was displaced by 
these observations, and acknowledge the dedicated efforts of the JCMT 
telescope operators for their valuable assistance with the observations.

We are grateful to Scott Barthelmy and Paul Butterworth for 
maintaining the GRB Coordinates Network (GCN), and to the other 
ground-based observers for the rapid dissemination of their 
burst results.

We acknowledge the data analysis facilities provided by the Starlink 
Project which is run by CCLRC on behalf of PPARC.

Some of the radio data referred to in this paper were drawn from the 
GRB Large Program at the VLA, http://www.vla.nrao.edu/astro/prop/largeprop/ .
The National Radio Astronomy Observatory is a facility of the National 
Science Foundation operated under cooperative agreement by Associated 
Universities, Inc.

The work at Rice University was supported in part by 
AFOSR/NSF grant number NSF AST-0123487.

\end{acknowledgements}


\begin{thebibliography}{}

\bibitem[2003]{scuba:ando03}
Ando, M., Ohta, K., Watanabe, C., et al. 2003, GCN 1884

\bibitem[2002]{scuba:arch02}
Archibald, E. N., Jenness, T., Holland, W. S., et al. 2002, MNRAS, 336, 1

\bibitem[2003]{scuba:barn03}
Barnard, V. E., Blain, A. W., Tanvir, N. R., et al. 2003, MNRAS, 338, 1

\bibitem[2004a]{scuba:bar04a}
Barnard, V., Schieven, G., Tilanus, R., et al. 2004a, GCN 2774

\bibitem[2004b]{scuba:bar04b}
Barnard, V., Schieven, G., Tilanus, R., et al. 2004b, GCN 2786

\bibitem[2000]{scuba:berger00}
Berger, E., Sari, R., \& Frail, D. A. 2000, ApJ, 545, 56

\bibitem[2002]{scuba:berg02}
Berger, E., Kulkarni, S. R., \& Frail, D. A. 2002, GCN 1612

\bibitem[2003]{scuba:berg03}
Berger, E., Cowie, L. L., Kulkarni, S. R., et al. 2003, ApJ, 588, 99

\bibitem[2003]{scuba:ber03}
Bersier, D., Stanek, K. Z., Winn, J. N., et al. 2003, ApJ, 584, L43

\bibitem[2003]{scuba:bert03}
Bertoldi, F., Frail, D. A., Berger, E., et al. 2003, GCN 1835
  
\bibitem[2004]{scuba:bgj04}
Bj\"ornsson, G., Gudmundsson, E. H., \& J\'ohannesson, G. 2004,
ApJ, 615, L77

\bibitem[2000]{scuba:bn00}
Blain, A. W., \& Natarajan, P. 2000, MNRAS, 312, L35

\bibitem[1999]{scuba:blain99}
Blain, A. W., Smail, I., Ivison, R. J., \& Kneib, J.-P. 1999, MNRAS, 302, 632

\bibitem[2002]{scuba:blain02}
Blain, A. W., Smail, I., Ivison, R. J., Kneib, J.-P., \& 
Frayer, D. T. 2002, \physrep, 369, 111

\bibitem[2003]{scuba:blain03}
Blain, A. W., Barnard, V. E., \& Chapman, S. C. 2003, MNRAS, 338, 733

\bibitem[1998]{scuba:bloom98}
Bloom, J. S., Frail, D. A., Kulkarni, S. R., et al., 1998, ApJ, 508, L21

\bibitem[2002]{scuba:bct02}
Bremer, M., \& Castro-Tirado, A. J. 2002, GCN 1590

\bibitem[2003]{scuba:but03}
Butler, N. R., Marshall, H. L., Ricker, G. R., et al. 2003, ApJ, 597, 1010

\bibitem[2004]{scuba:but04}
Butler, N., Vanderspek, R., Marshall, H. L., et al. 2004, GCN 2808

\bibitem[1999]{scuba:cy99}
Carilli, C. L., \& Yun, M. S. 1999, ApJ, 513, L13

\bibitem[2003]{scuba:ct03}
Castro-Tirado, A. J., Mateo Sanguino, T. J., de Ugarte Postigo, A.,
et al. 2003, GCN 1826

\bibitem[2003]{scuba:chap03}
Chapman, S. C., Blain, A. W., Ivison, R. J., \& Smail, I. R. 2003,
Nature, 422, 695

\bibitem[2004]{scuba:chap04}
Chapman, S. C., Smail, I., Blain, A. W., \& Ivison, R. J. 
2004, ApJ, 614, 671

\bibitem[2005]{scuba:chap05}
Chapman, S. C., Blain, A. W., Smail, I., \& Ivison, R. J. 2005, ApJ, in press
(astro-ph/0412573)

\bibitem[2000]{scuba:cl00}
Chevalier, R. A., \& Li, Z.-Y. 2000, ApJ, 536, 195

\bibitem[2003]{scuba:cf03}
Chornock, R., \& Filippenko, A. V. 2003, GCN 1897

\bibitem[2004]{scuba:christ04}
Christensen, L., Hjorth, J., \& Gorosabel, J. 2004, A\&A, 425, 913

\bibitem[1992]{scuba:condon}
Condon, J. J. 1992, ARAA, 30, 575

\bibitem[2003]{scuba:con03}
Conselice, C. J., Chapman, S. C., \& Windhorst, R. A. 2003, ApJ, 596, L5

\bibitem[2004]{scuba:cov04}
Covino, S., Malesani, D., Tagliaferri, G., et al. 2004, GCN 2803

\bibitem[2003]{scuba:crew03}
Crew, G. B., Lamb, D. Q., Ricker, G. R., et al. 2003, ApJ, 599, 387

\bibitem[2004]{scuba:dcnp04}
da Costa, G., Noel, N., \& Price, P. A. 2004, GCN 2765

\bibitem[2004]{scuba:dav04}
D'Avanzo, P., Covino, S., Antonelli, L. A., et al. 2004, GCN 2788

\bibitem[2003]{scuba:dv03}
Della Valle, M., Malesani, D., Benetti, S., et al. 2003, A\&A, 406, L33

\bibitem[2002]{scuba:djo02}
Djorgovski, S. G., Barth, A., Price, P., et al. 2002, GCN 1620

\bibitem[2004]{scuba:dull04}
Dullighan, A., Ricker, G., Butler, N., \& Vanderspek, R. 2004,
in AIP Conf. Proc. 727, Gamma-Ray Bursts: 30 Years of Discovery: 
Gamma-Ray Burst Symposium, eds. E. E. Fenimore \& M. Galassi, 467

\bibitem[2004]{scuba:dun04}
Dunlop, J. S., McLure, R. J., Yamada, T., et al. 2004, MNRAS, 350, 769

\bibitem[2003a]{scuba:fox03a}
Fox, D. W., Yost, S., Kulkarni, S. R., et al. 2003a, Nature, 422, 284

\bibitem[2003b]{scuba:fox03b}
Fox, D. W., Price, P. A., Soderberg, A. M. et al. 2003b, ApJ, 586, L5

\bibitem[2003c]{scuba:fox03c}
Fox, D. W., Chen, H. W., \& Price, P. A. 2003c, GCN 1879

\bibitem[2002]{scuba:fb02}
Frail, D. A., \& Berger, E. 2002, GCN 1574

\bibitem[2003]{scuba:fb03}
Frail, D. A., \& Berger, E. 2003, GCN 1827

\bibitem[2000]{scuba:frail00}
Frail, D. A., Berger, E., Galama, T., et al. 2000, ApJ, 538, L129

\bibitem[2002]{scuba:frail02}
Frail, D. A., Bertoldi, F., Moriarty-Schieven, G. H., et al. 
2002, ApJ, 565, 829

\bibitem[2003a]{scuba:frail03a}
Frail, D. A., Yost, S. A., Berger, E., et al. 2003a, ApJ, 590, 992

\bibitem[2003b]{scuba:frail03b}
Frail, D. A., Kulkarni, S. R., Berger, E., \& Wieringa, M. H. 2003b,
ApJ, 125, 2299

\bibitem[2004]{scuba:fug04}
Fugazza, D., Fiore, F., Covino, S. et al. 2004, GCN 2782

\bibitem[2003]{scuba:fynbo03}
Fynbo, J. P. U., Jakobsson, P., M{\o}ller, P., et al. 2003, A\&A, 406, L63

\bibitem[2004]{scuba:fyn04}
Fynbo, J. P. U., Jensen, B. L., Pedersen, K., et al. 2004, GCN 2802

\bibitem[1998]{scuba:galama98}
Galama, T. J., Wijers, R. A. M. J., Bremer, M., et al. 1998, ApJ, 500, L97

\bibitem[1999]{scuba:gal99nat}
Galama, T. J., Briggs, M. S., Wijers, R. A. M. J., et al. 1999, 
Nature, 398, 394

\bibitem[2004]{scuba:gal04}
Galassi, M., Ricker, G., Atteia, J-L., et al. 2004, GCN 2770

\bibitem[2003]{scuba:gar03}
Garnavich, P. 2003, GCN 1848

\bibitem[2004]{scuba:swift04}
Gehrels, N., Chincarini, G., Giommi, P., et al. 2004, ApJ, 611, 1005

\bibitem[2002]{scuba:gs02}
Granot, J., \& Sari, R. 2002, ApJ, 568, 820

\bibitem[2000]{scuba:gps00}
Granot, J., Piran, T., \& Sari, R. 2000, ApJ, 534, L163

\bibitem[2003]{scuba:gnp03}
Granot, J., Nakar, E., \& Piran, T. 2003, Nature, 426, 138

\bibitem[2001]{scuba:har01}
Harrison, F. A., Yost, S. A., Sari, R., et al. 2001, ApJ, 559, 123

\bibitem[2002]{scuba:hoge02}
Hoge, J., Willott, C., Grimes, J., Tilanus, R., \&
Moriarty-Schieven, G. 2002, GCN 1742

\bibitem[2003]{scuba:hoge03}
Hoge, J. C., Stevens, J. A., Moriarty-Schieven, G., \&
Tilanus, R. P. J. 2003, GCN 1832

\bibitem[2003]{scuba:hol03}
Holland, S. T., Weidinger, M., Fynbo, J. P. U., et al. 2003, AJ, 125, 2291

\bibitem[2004]{scuba:hol04}
Holland, S. T., Bersier, D., Bloom, J. S., et al. 2004, AJ, 128, 1955

\bibitem[1999]{scuba:hol99}
Holland, W. S., Robson, E. I., Gear, W. K., et al. 1999, MNRAS, 303, 659

\bibitem[2005]{scuba:ioc05}
Inoue, S., Omukai, K., \& Ciardi, B. 2005, MNRAS, submitted
(astro-ph/0502218)

\bibitem[2000]{scuba:jen00}
Jenness, T., \& Lightfoot, J. F. 2000, Starlink User Note 216, 
Starlink Project, CLRC

\bibitem[2004]{scuba:kah04}
Kahharov, B., Asfandiyarov, I., Ibrahimov, M., et al. 2004, GCN 2775

\bibitem[2003]{scuba:kato03}
Kato, D., Nagata, T., \& Kawai, N. 2003, GCN 1830

\bibitem[2003]{scuba:kawai03}
Kawai, N., Ricker, G., Atteia, J-L, et al. 2003, GCN 1816

\bibitem[2002]{scuba:kemp02}
Kemp, J., Fiege, J., Coppin, K., et al. 2002, GCN 1619

\bibitem[2004]{scuba:klose04}
Klose, S., Greiner, J., Rau, A., et al. 2004, AJ, 128, 1942

\bibitem[1993]{scuba:kouv93}
Kouveliotou, C., Meegan, C. A., Fishman, G. J., et al. 1993, ApJ, 413, L101

\bibitem[1999]{scuba:kfs99}
Kulkarni, S. R., Frail, D. A., Sari, R., et al. 1999, ApJ, 522, L97

\bibitem[2002]{scuba:lamb02}
Lamb, D., Ricker, G., Atteia, J-L, et al. 2002, GCN 1600

\bibitem[2003]{scuba:laz03}
Lazzati, D., Covino, S., di Serego Alighieri, S., et al. 
2003, A\&A, 410, 823

\bibitem[2003]{scuba:levan03}
Levan, A., Merrill, M., Rol, E., et al. 2003, GCN 1818

\bibitem[2005]{scuba:levan05}
Levan, A., et al. 2005, in press

\bibitem[2003]{scuba:li03}
Li, W., Filippenko, A. V., Chornock, R., \& Jha, S. 2003, ApJ, 586, L9

\bibitem[2003]{scuba:lc03}
Li, Z-Y, \& Chevalier, R. A. 2003, ApJ, 589, L69

\bibitem[2000]{scuba:lw00}
Livio, M., \& Waxman, E. 2000, ApJ, 538, 187

\bibitem[2004]{scuba:mae04}
Maeno, S., Sonoda, E., Matsuo, Y., \& Yamauchi, M. 2004, GCN 2772

\bibitem[2000]{scuba:mas00}
Masetti, N., Bartolini, C., Bernabei, S., et al. 2000, A\&A, 359, L23

\bibitem[2003]{scuba:mas03}
Masetti, N., Palazzi, E., Pian, E., et al. 2003, GCN 1823

\bibitem[2003]{scuba:mat03}
Matheson, T., Garnavich, P. M., Foltz, C., et al. 2003, ApJ, 582, L5

\bibitem[2002]{scuba:mes02}
M\'esz\'aros, P. 2002, ARA\&A, 40, 137

\bibitem[2003]{scuba:mir03}
Mirabal, N., Halpern, J. P., Chornock, R., et al. 2003, ApJ, 595, 935

\bibitem[2000]{scuba:pk00}
Panaitescu, A., \& Kumar, P. 2000, ApJ, 543, 66

\bibitem[2004]{scuba:pk04}
Panaitescu, A., \& Kumar, P. 2004, MNRAS, 350, 213

\bibitem[2001]{scuba:pkn01}
Panaitescu, A., Kumar, P., \& Narayan, R. 2001, ApJ, 561, L171

\bibitem[2003]{scuba:pan03}
Pandey, S. B., Anupama, G. C., Sagar, R., et al. 2003, A\&A, 408, L21

\bibitem[2004]{scuba:pan04}
Pandey, S. B., Sagar, R., Anupama, G. C., et al. 2004, A\&A, 417, 919

\bibitem[2002]{scuba:pwb02}
Park, H. S., Williams, G., \& Barthelmy, S. 2002, GCN 1736

\bibitem[2003]{scuba:ped03}
Pedersen, K., Fynbo, J., Hjorth, J., et al. 2003, GCN 1924

\bibitem[1999]{scuba:piran99}
Piran, T. 1999, Physics Reports, 314, 575

\bibitem[2002]{scuba:pool02}
Pooley, G. 2002, GCN 1604

\bibitem[2002]{scuba:price02}
Price, P. A., Berger, E., Kulkarni, S. R., et al. 2002, ApJ, 573, 85

\bibitem[2003a]{scuba:price03a}
Price, P. A., Fox, D. W., \& Chen, H. W. 2003a, GCN 1880

\bibitem[2003b]{scuba:price03b}
Price, P. A., Fox, D. W., \& Djorgovski, S. G., et al. 2003b, GCN 1889

\bibitem[2004a]{scuba:pdcn04a}
Price, P. A., da Costa, G., \& Noel, N. 2004a, GCN 2766

\bibitem[2004b]{scuba:pdcn04b}
Price, P. A., da Costa, G., \& Noel, N. 2004b, GCN 2771

\bibitem[2004c]{scuba:prr04}
Price, P. A., Roth, K., Rich, J., et al. 2004c, GCN 2791

\bibitem[2002]{scuba:rtb02}
Ramirez-Ruiz, E., Trentham, N., \& Blain, A. W. 2002, MNRAS, 329, 465

\bibitem[2002]{scuba:rs02}
Rol, E., \& Strom, R. 2002, GCN 1777

\bibitem[2003]{scuba:rw03}
Rol, E., \& Wijers, R. 2003, GCN 1867

\bibitem[2003]{scuba:rol03}
Rol, E., Wijers, R. A. M. J., Fynbo, J. P. U., et al. 2003, A\&A, 405, L23

\bibitem[2000]{scuba:sm00}
Sari, R., \& M\'esz\'aros, P. 2000, ApJ, 535, L33

\bibitem[1998]{scuba:spn98}
Sari, R., Piran, T., \& Narayan, R. 1998, ApJ, 497, L17

\bibitem[2003]{scuba:sgh03}
Schaefer, B. E., Gerardy, C. L., H\"oflich, P., et al. 2003, ApJ, 588, 387

\bibitem[2002]{scuba:shir02}
Shirasaki, Y., Graziani, C., Matsuoka, M., et al. 2002, GCN 1565

\bibitem[1999]{scuba:smith99}
Smith, I. A., Tilanus, R. P. J., Van Paradijs, J., et al. 1999, A\&A, 347, 92

\bibitem[2000]{scuba:smith00}
Smith, I. A., Van Paradijs, J., Tilanus, R. P. J., et al. 2000, 
in Gamma-Ray Bursts: 5th Huntsville Symposium,
ed. R. M. Kippen, R. S. Mallozzi, \& G. J. Fishman (New York: AIP), 326

\bibitem[2001]{scuba:smith01}
Smith, I. A., Tilanus, R. P. J., Wijers, R. A. M. J., et al. 2001, A\&A, 
380, 81

\bibitem[2005]{scuba:smith05}
Smith, I. A., Tilanus, R. P. J., Tanvir, N., et al. 2005, A\&A, submitted

\bibitem[2004]{scuba:sod04}
Soderberg, A. M., \& Frail, D. A. 2004, GCN 2787

\bibitem[2003]{scuba:suz03}
Suzuki, M., Shirasaki, Y., Graziani, C., et al. 2003, GCN 1888

\bibitem[2004]{scuba:tan04}
Tanvir, N. R., Barnard, V. E., Blain, A. W., et al. 2004, MNRAS, 352, 1073

\bibitem[2002]{scuba:trent02}
Trentham, N., Ramirez-Ruiz, E., \& Blain, A. W. 2002, MNRAS, 334, 983

\bibitem[2003]{scuba:ukiy03}
Uemura, M., Kato, T., Ishioka, R., \& Yamaoka, H. 2003, PASJ, 55, L31

\bibitem[2000]{scuba:vanp00}
Van Paradijs, J., Kouveliotou, C., \& Wijers, R. A. M. J. 2000, ARA\&A, 
38, 379

\bibitem[2003]{scuba:vrba03}
Vrba, F., Luginbuhl, C., \& Henden, A. 2003, GCN 1822

\bibitem[2003]{scuba:vrees03}
Vreeswijk, P., Fruchter, A., Hjorth, J., \& Kouveliotou, C. 2002, GCN 1785

\bibitem[1999]{scuba:wg99}
Wijers, R. A. M. J., \& Galama, T. J. 1999, ApJ, 523, 177

\bibitem[2002]{scuba:wout02}
Wouterloot, J., Davis, G., Naylor, D., Tilanus, R., 
\& Moriarty-Schieven, G. 2002, GCN 1627

\bibitem[2002]{scuba:woz02}
Wozniak, P., Vestrand, W. T., Starr, D., et al. 2002, GCN 1757

\bibitem[2002]{scuba:yost02}
Yost, S. A., Frail, D. A., Harrison, F. A., et al. 2002, ApJ, 577, 155

\bibitem[2003]{scuba:yost03}
Yost, S. A., Harrison, F. A., Sari, R., \& Frail, D. A. 2003, ApJ, 597, 459

\end{thebibliography}
\end{document}